\documentstyle[preprint,eqsecnum,aps]{revtex}
\begin{document}
\draft
\preprint{SNUTP 98-014 
        \hspace{-28.5mm}\raisebox{2.4ex} {SOGANG-HEP 230/97}}
\title{Non-Abelian Ramond-Neveu-Schwarz String Theory}
\author{Hyun Seok Yang, Inbo Kim, and Bum-Hoon Lee}
\address{Department of Physics, Sogang University, Seoul 121-742, Korea}
\maketitle

\begin{abstract}
We construct a locally supersymmetric worldsheet formulation of a
non-Abelian Ramond-Neveu-Schwarz (NARNS) string theory where the string
coordinates are noncommuting matrices in a group $U(N)$. This is
described by the two dimensional supergravity coupled to 
supersymmetric Yang-Mills
fields and adjoint matters in the gauge group $U(N)$. 
We show that our NARNS string theory has a free
string limit where 
it becomes N-copies of usual RNS string which can be described by 
the orbifold conformal field theory corresponding to the 
covariant worldsheet version of the Matrix string theory of 
Dijkgraaf, Verlinde and Verlinde. 
In the weak coupling limit, i.e. $g_s \rightarrow 0$ where $g_s$ is
the coupling constant of our theory related with the Yang-Mills coupling
as $g_{YM}^{-2}=\alpha' g_s^2$, a new additional dimension
appears in the string spectrum and it can be speculatively interpreted
as the compactified {\it eleven} dimensional coordinate whose dynamics
is given by an orbifold O(N) sigma model.
\end{abstract}

\def\be{\begin{equation}}
\def\ee{\end{equation}}
\def\bea{\begin{eqnarray}}
\def\eea{\end{eqnarray}}
\def\ba{\begin{array}}
\def\ea{\end{array}}
\def\l{\label}
\def\r{\ref}
\def\c{\cite}
\def\n{\nonumber}
\narrowtext

\section{Introduction and Motivation}

The nonperturbative formulation of string theory needs a mysterious
eleven dimensional M-theory \c{HT95,Witt95}, 
which is a strong coupling limit of type
IIA superstring theory and that its low energy limit is eleven dimensional 
supergravity which has membrane and M5-brane 
as fundamental degree of freedom as
well as graviton. Though one is lacking an intrinsic definition of
M-theory in terms of its underlying degrees of freedom, its mere
existence led to many powerful predictions or simplifications of
superstring duality \c{Schw96}. 
Major step forward was taken by Banks, Fischler,
Shenker and Susskind (BFSS) whose conjecture is that M-theory quantum
dynamics in the infinite momentum frame (IMF) is described by $U(N)$
supersymmetric Yang-Mills (SYM) quantum mechanics \c{BFSS}. 
One of the remarkable
pictures of the BFSS matrix theory is that the spacetime coordinates
live in a linear space of matrices which define so-called
noncommutative spacetime geometry \c{Witt96}. 
The classical commutative geometry
is only sensible concept in a long distance regime. The matrix theory
provides a natural and simple mechanism for the appearance of a
noncommutative geometry at short distances \c{Matrix}.

It has been shown that the BFSS matrix theory compactified on a tiny 
circle, Matrix string theory (MST), provides a nonperturbative 
definition of the weakly coupled type IIA string theory 
\c{Motl,DVV}. The string
coordinates of the MST are also matrices taking values in the
non-Abelian gauge group although they become usual commutative
C-numbers in the weak coupling limit which corresponds to the zero
size limit of the compactified circle. The beautiful picture arises in
the MST. It has a description of the Hilbert space of {\it second}
quantized string theory \c{DVV}. The second quantized string is due to the
residual discrete Weyl symmetry of the gauge group acting on the
matrix elements within the Cartan subalgebra. In the BFSS matrix
theory, this Weyl symmetry gives the conventional spin statistics on
the states of the D0-brane Fock space \c{BFSS}. 
  
Our motivation to construct a non-Abelian Ramond-Neveu-Schwarz 
(NARNS) string theory where the string coordinates are noncommuting 
matrices comes from this new picture, ``noncommutative spacetime
geometry''.  We think that there must exist a Lorentz
invariant worldsheet formulation of MST which is a Green-Schwarz 
\c{GS} light-cone formulation. What is the NARNS string theory and 
how the NARNS string theory should be constructed? It is a
generalization of the usual RNS string theory 
\c{RNS} in the way that the
string coordinates are noncommuting matrices in a group $G$, which
depends on the worldsheet coordinate. Therefore the NARNS string
theory is the two dimensional (2D) supergravity theory coupled to 
SYM fields and adjoint matters (string
coordinates) in the gauge group $G$.    

Recently the noncommutative spacetime 
picture of string theory appeared in an interesting way as 
the form of spacetime uncertainty relation by Yoneya and Li \c{yoneya}. 
They argue that the spacetime uncertainty relation of the form 
$\Delta X \Delta T \ge \alpha'$ for the observability of the distances
with respect to time, $\Delta T$ and space, $\Delta X$, is 
universally valid in string theory including nonperturbative objects,
D-branes \c{tasi} and this relation can be derived as a direct consequence of
the worldsheet conformal invariance. It implies that the fumdamental
constant $\alpha'$ of Nature representing the string size has a 
fundamental significance as the constant $c$ (Lorentz covariance) and
$\hbar$ (quantum mechanics). If we innocently accept their argument,
the spacetime probed by string should be a noncommutative object for
short distances compared to the string scale $l_s \equiv \sqrt
{\alpha'}$. Only for $\alpha' \rightarrow 0$ limit, the classical
geometry appears. This is an another motivation for our NARNS string theory. 

The organization of this paper is as follows. 
In Sec.II, we formulate the superspace of 2D 
supergravity coupled to SYM fields 
for the purpose of constructing a locally supersymmetric worldsheet 
formulation of the NARNS string theory described above. Our
superspace formulation of $N=1$ SYM theory coupled to 2D
supergravity is a new one up to our knowledge. In Sec.III, we
explicitly construct the NARNS string theory and show  
that it has a free
string limit where it becomes N-copies of usual RNS string. 
And we observe that, 
in the weak coupling limit, i.e. $g_s \rightarrow 0$ where $g_s$ is
the coupling constant of our theory related with the Yang-Mills coupling
as $g_{YM}^{-2}=\alpha' g_s^2$, a new additional dimension
appears in the string spectrum and it can be speculatively interpreted
as the compactified {\it eleven} dimensional coordinate. 
In Sec.IV, we argue that 
the NARNS string theory is described by the orbifold 
conformal field theory (CFT) \c{orbifold}, 
essentially second quantized string theory, 
contrary to the ordinary RNS string which 
has a first quantized description. 
In the $g_s \ll 1$ limit with fixed $\alpha'$, 
the SYM part must be considered. 
Nevertheless, the full superconformal symmetry of the NARNS 
string theory is preserved in a particular configuration. 
In the limit, the additional degree of freedom interpreted as compactified
eleven dimension in this paper is
interestingly described by the orbifold O(N) sigma model predicting that 
the size of this dimension increases in the ultraviolet limit 
and decreases in the infrared limit \c{on}. 
In Sec.V, we discuss many aspects
of NARNS string theory. In Appendices, our conventions and some
identities are listed and the details of the superspace
formulation of 2D SYM theory coupled 
to supergravity are presented.

\section{2D Supergravity Coupled to Super-Yang-Mills Theory}

In this section, we will formulate the superspace of 2D 
supergravity coupled to SYM fields 
for the purpose of constructing a locally supersymmetric worldsheet 
formulation of a NARNS string
theory where the string coordinates are noncommuting matrices 
in a compact Lie group $G$. 
We will closely follow the Wess and Bagger \c{wb} on the superspace and 
Howe \c{howe} on the 2D supergravity.

The 2D superspace has two kinds of supercoordinate
indices, a curved index $M=(m,n;\mu, \nu)$ and a tangent 
index $A=(a,b;\alpha, \beta)$. The coordinates of superspace, 
$z^M=(\sigma^m, \theta^\mu)$, obey the following multiplication law:          
\be
\l{csc}
z^M z^N=(-)^{mn} z^N z^M.
\ee
At each point in superspace the one-form basis $E^A(z)$ define 
a local reference frame:
\be
\l{sz}
E^A=dz^M E_M\,^A(z),
\ee
where superzweibein $E_M\,^A(z)$ is an arbitrary invertible superfield,
\bea
\l{eie}
E_M\,^A(z)E_A\,^N(z)&=&\delta_M\,^N\n\\
E_A\,^M(z)E_M\,^B(z)&=&\delta_A\,^B.
\eea
The exterior derivative may be written in terms of the differential
operator in the local frame (\r{sz})
\be
\l{exd}
d=dz^M\frac{\partial}{\partial z^M}=E^AD_A,
\ee
where
\[ D_A=E_A^N \frac{\partial}{\partial z^N}.\]

The tangent frame $E^A(z)$ is locally Lorentz covariant:
\be
\l{lte}
\delta E^A(Z)=E^B L_B\,^A(z),
\ee
where the Lorentz generators $L_B\,^A$ have two irreducible components:
\bea
\l{lg}
&& L_B\,^A(z)=L(z){\cal E}_B\,^A\n\\
&& {\cal E}_B\,^A = \pmatrix{
              -\epsilon_a\,^b & 0\cr
               0 & \frac{1}{2}(\gamma_5)_{\alpha}\,^{\beta}\cr}.
\eea
To define covariant derivative on 2D Lorentz group we must introduce 
a connection form
\be
\l{co}
\Omega_B\,^A=\Omega {\cal E}_B\,^A=dz^M \Omega_{M,B} \,^A,
\ee
transforming as follows under the Lorentz group:
\be
\l{lts}
\delta \Omega=-dL.
\ee
The connections allow us the covariant derivatives, for example, 
for a one-form $V$,
\be
\l{locovd}
{\cal D}V=dV+V\Omega.
\ee
Then the torsion two-form, $T^A$, is defined as the covariant 
derivative of the vielbein and the curvature two-form, $R_A\,^B$, 
in terms of the connection:
\bea
\l{tc}
T^A &=&{\cal D}E^A=dE^A+E^B\Omega_B\,^A\n\\
   &=& \frac{1}{2}E^C E^B T^A_{BC},\\
R_A\,^B &=& d\Omega_A\,^B +\Omega_A\,^C \Omega_C\,^B \n\\
        &=& \frac{1}{2}E^D E^C R_{CD,A}\,^B.
\eea
They satisfy the Bianchi identies,
\bea
\l{bi}
&&{\cal D}T^A=E^B R_B\,^A\cr
&&{\cal D}R_B\,^A=0.
\eea
The 2D Lorentz group structure, Eq.(\r{lg}), 
allows us to make the simplication
\[R_A\,^B=F {\cal E}_A\,^B ; \;\;\; F=d\Omega.\]

In order to formulate 2D supergravity coupled to $N=1$
SYM theory, we must introduce a Lie algebra valued
one-form:
\bea
\l{gaugep}
&& A=dz^M A_M = E^B A_B,\\
&& A_B=A_B^r T^r,\;\;\;r=1,\cdots,\mbox{dim}G,\n
\eea
where the matrices $T$ are the hermitean generators of the structure
group $G$. Under a local structure group represented by $U=e^{iX}$
where the gauge parameter $X$ is a scalar superfield, the gauge
connections transform as 
\be
\l{gaugetr}
A'=U^{-1}AU-iU^{-1}dU
\ee
and we can define gauge covariant derivatives as before
\be
\l{gcovd}
\nabla\phi = d\phi -ig_{YM}( \phi A \pm A\phi)
\ee
for a superfield $\phi$ in the adjoint representation of the group $G$. 
Here $+$ is for odd form $\phi$ and $-$ for even form. 
The curvature two-form tensor which can be constructed from the
connection and its derivatives is defined as follows 
\bea
\l{fiels}
F&=&\frac{1}{2}dz^M dz^N F_{NM}=\frac{1}{2}E^AE^B F_{BA}\n\\
 &=&dA-ig_{YM}A^2,
\eea
which covariantly transforms under the local structure group
\be
\l{gaugetrf}
F'=U^{-1}FU.
\ee
The curvature tensor, ``field strength'', may be then read as 
the component form
\be
\l{comfs}
F_{BC}={\cal D}_BA_C-(-)^{bc}{\cal D}_CA_B+ig_{YM}(A_BA_C-(-)^{bc}A_CA_B)
+T_{BC}^DA_D.
\ee
As in the Eq.(\r{bi}), the field strengths introduced above similary 
satisfy Bianchi identities by virtue of their definition in terms 
of `potentials':
\be
\l{gbi}
\nabla F=dF-ig_{YM}(FA-AF)=0,
\ee
in component form, which are
\be
\l{gbic}
\Delta_{[A}F_{BC\}}+T_{[AB}\;^{|D}F_{D|C\}}=0,
\ee
where $[\;\}$ represents graded antisymmetrization and 
the derivatives $\Delta_A F_{BC}=\nabla_AF_{BC}-\Omega_{A,B}\;^DF_{DC}
-\Omega_{A,C}\;^DF_{BD}$ are the gauge and superspace 
covariant derivatives.
 
As the four dimensional case, we take the proper constraints on
supertorsion to reduce the number of component fields, which must be
Lorentz covariant, gauge covariant, and supersymmetric and should not
restrict the $\sigma$-dependence of the component fields \c{wb}:
\be
\l{ct}
 T_{\alpha\beta}^{a}=2i(\gamma^a)_{\alpha\beta},
\;\;\;T_{\beta\gamma}^{\alpha}=T_{bc}^{a}=0.
\ee
One can solve the Bianchi identities (\r{bi}) subject to the
constraints (\r{ct})\footnote{In fact, it can
be showed that, for the choice of the tangent space group made here, 
the second Bianchi identity is derived from the first 
through its covariant derivatives, so that it is
sufficient only to solve the first one.} and find that 
all the components of the curvature 
and the torsion can be expressed in terms of one scalar superfield $S$
\c{howe}. 
Thus the supergeometry can be determined by the Bianchi identities 
and the torsion constraints.

In order to eliminate the gauge degrees of freedom coming from 
the super-reparameterization and local Lorentz transformation, 
we impose the proper conditions on the superzweibein and the
connection, so-called, Wess-Zumino gauge \c{wb}. 
The component fields can then be derived from the superfield 
language in terms of their $\theta$ expansions \c{howe};
\bea
\l{come}
&&{E_m}^a={e_m}^a+i{\bar \theta}\gamma^a\chi_m
+\frac{i}{4}{\bar \theta}\theta {e_m}^a A,\cr
&&{E_m}^{\alpha}=\frac{1}{2}{\chi_m}^{\alpha}+
\frac{1}{2}\theta^\mu (\gamma_5)_\mu\,^\alpha 
\omega_m-\frac{1}{4}\theta^\mu (\gamma_m)_\mu\,^\alpha A
+\frac{i}{2}{\bar \theta}\theta (-\frac{3}{8}{\chi_m}^{\alpha}A
+\frac{i}{4}(\gamma_m)^{\alpha\beta}\psi_\beta),\\
&&{E_\mu}^a=i\theta^{\lambda}(\gamma^a)_{\lambda\mu},\cr
&&{E_\mu}^{\alpha}=\delta_\mu\,^\alpha(1-\frac{i}{8}{\bar \theta}\theta A),\n
\eea
and
\bea
\l{spinconn}
&&\Omega_m=\omega_m-\frac{i}{2}{\bar \theta}\gamma_5\chi_m A-\frac{1}{2}
{\bar \theta}\gamma_5\gamma_m\psi
-\frac{i}{4}{\bar \theta}\theta (\omega_m A+
e\epsilon_m\,^n\partial_n A+
\frac{1}{2}{\bar \chi}_n\gamma_5\gamma_m \gamma^n\psi),\cr
&&\Omega_\mu=-\frac{i}{2}\theta^\lambda(\gamma_5)_{\lambda\mu}A,\\
&&\omega_m=\frac{1}{e}e_{ma}\epsilon^{nl}\partial_n e_l\,^a
+\frac{i}{2}{\bar \chi}_m\gamma_5\gamma^n \chi_n,\cr
&&\psi=\frac{2i}{e}\epsilon^{mn}\gamma_5(\partial_m \chi_n
-\frac{1}{2}\omega_m\gamma_5 \chi_n)-\frac{i}{2}\gamma^m\chi_m A.\n 
\eea
Here $A$ and $\psi$ are the first and second components of 
the curvature superfield $S$,
\bea
\l{curv}
&&S=A+{\bar \theta}\psi+ \frac{i}{2}{\bar \theta}\theta C \n\\
&&C=-\frac{2}{e}\epsilon^{mn}\partial_m \omega_n-
\frac{i}{4e}\epsilon^{mn}{\bar \chi}_m\gamma_5 \chi_n A-
\frac{1}{2}{\bar \chi}_m\gamma^m \psi-\frac{1}{2}A^2,
\eea
and $e=|\mbox{det}\, {e_m}^a|$. 
The superdeterminant must be introduced to construct invariant actions
from superfields and can be calculated from superzweibeins, Eq.(\r{come})
\be
\l{supd}
E=sdetE_M\,^A=e(1+\frac{i}{2}{\bar \theta}\gamma^m\chi_m+
\frac{i}{4}{\bar \theta}\theta A-
\frac{1}{8}{\bar \theta}\theta\frac{\epsilon^{mn}}{e}
{\bar \chi}_m\gamma_5 \chi_n).
\ee

The superspace transformations are constructed from the 
super-reparameterization and local Lorentz transformations of superspace. 
The infinitesimal changes in the vielbein and the connection 
under the superspace transformations are given by
\bea
\l{steo}
&& \delta E_M\,^A=\xi^N\partial_N E_M\,^A+\partial_M \xi^N 
E_N\,^A+ E_M\,^B L_B\,^A \cr
&& \delta \Omega_M=\xi^N\partial_N \Omega_M+
\partial_M \xi^N \Omega_N-\partial_M L.
\eea
From the superspace transformations of $E_\mu\,^A$ and $\Omega_\mu$ 
one finds that the transformation parameters $\xi^M$ and 
$L$ can be decomposed as the following forms \c{howe}
\bea
\l{tpxi}
&&\xi^m=f^m-i{\bar \alpha}\gamma^m \theta +\frac{1}{4}{\bar \theta}\theta 
{\bar \alpha}\gamma^n\gamma^m\chi_n,\cr
&&\xi^\mu=\alpha^\mu-\frac{1}{2}\theta^\nu (\gamma_5)_\nu\,^\mu l
+\frac{i}{2}{\bar \alpha}\gamma^m \theta {\chi_m}^\mu
-\frac{1}{8}{\bar \theta}\theta 
{\bar \alpha}\gamma^n\gamma^m\chi_n{\chi_m}^\mu
+\frac{i}{4}{\bar \theta}\theta
({\bar \alpha}\gamma_5\gamma^m)^\mu \omega_m,\\
&&L=l- \frac{i}{2}{\bar \alpha}\gamma_5\theta A-
i{\bar \alpha}\gamma^m\theta \omega_m+
\frac{1}{8}{\bar \theta}\theta {\bar \alpha}\gamma^m\gamma_5\chi_m A
+\frac{i}{4}{\bar \theta}\theta {\bar \alpha}\gamma_5\psi
+\frac{1}{4}{\bar \theta}\theta{\bar \alpha}\gamma^n\gamma^m\chi_n\omega_m,\n
\eea 
where $(f, \alpha, l)$ correspond to coordinate, local supersymmetry 
and Lorentz transformations respectively. 
The supersymmetry transformations for the supergravity multiplet 
$({e_m}^a, \chi_m, A)$ may be read off and one finds
\be
\l{stgm}
\delta {e_m}^a=i{\bar \alpha}\gamma^a\chi_m,\;\;\delta \chi_m=
2{\cal D}_m\alpha
+\frac{1}{2}\gamma_m\alpha A,\;\;\delta A={\bar \alpha}\psi,
\ee
where ${\cal
D}_m\alpha=\partial_m\alpha-\frac{1}{2}\omega_m\gamma_5\alpha$ 
is the covariant derivative of the spinor $\alpha$.

In order to formulate the 2D SYM 
theory in superspace \c{ht,sakai}, we need some constraints eliminating 
superflous components in the superfields $A_B(z)$, which are analogue
of the supertorsion constraints (\r{ct}) in the supergravity sector. 
According to the definite
analogy between 2D and 4D gauge theory, 
one may choose the constraints as $F_{\alpha\beta}=0$. 
However we will find that the Bianchi identity (\r{gbi}) 
together with these constraints makes the 2D 
SYM theory trivial. In order to construct 
an appropriate nontrivial superspace, we instead impose 
the alternative constraints on the theory as follows:
\be
\l{cf}
(\gamma_a)^{\alpha\beta} F_{\alpha\beta}=0. 
\ee
Note that the above constraints can be solved by introducing 
the scalar superfield $W$
\be
\l{sfs}
 F_{\alpha\beta}=(\gamma_5)_{\alpha\beta}W(z).
\ee
Then the Eq.(\r{comfs}) implies that the scalar field strength $W(z)$
is represented in terms of the spinor potentials:
\be
\l{sfse}
W(z)=-(\gamma_5)^{\alpha\beta}({\cal D}_\alpha A_\beta 
+ ig_{YM}A_\alpha A_\beta).
\ee 

Now one can similarly solve the Bianchi
identity (\r{gbi}) or (\r{gbic}) subject to the constraints (\r{cf}) 
and find that the Yang-Mills mutiplet can be expressed in terms of 
one spinor superfield $A_{\alpha}(z)$.\footnote{In fact, it is not
necessary to solve the Bianchi identities because, by solving 
the constraints (\r{cf}) directly, we can easily determine 
the vector potential $A_m(z)$ in terms of 
spinor potential $A_\alpha(z)$.} 
The detailed solutions of
the Yang-Mills Bianchi identity and the superspace formulation of 
2D SYM theory coupled to supergravity 
will be presented to the Appendix B.

The spinor superfields $A_{\alpha}(z)$ correspond to the Yang-Mills
vector multiplet in the adjoint representation of the gauge group
$U(N)$ and are expanded as
\be
\l{vecmul}
A_{\alpha}(z)=\eta_{\alpha}(\sigma)+i(\gamma^m\theta)_{\alpha} 
g_m(\sigma)+\theta_{\alpha}n(\sigma)+i(\gamma_5\theta)_{\alpha} 
\phi(\sigma)+\frac{1}{2}{\bar \theta}\theta b_{\alpha}(\sigma),
\ee
where
\bea
&& g_m(\sigma)=v_m-\frac{1}{2}{\bar \chi}_m\eta,\n\\
&& b_{\alpha}(\sigma)=(2\lambda-\frac{1}{2}\gamma^m\gamma^n\chi_m g_n
-i\gamma^m\Delta_m \eta +\frac{i}{2}\gamma^m\chi_m n
-\frac{i}{4}A\eta+g_{YM}\gamma_5[\phi,\eta])_{\alpha},\n
\eea
where $\Delta_m \eta =\partial_m \eta -\frac{1}{2}\omega_m\gamma_5
\eta +ig_{YM}[v_m, \eta]$. 
The gauge transformation on $A_{\alpha}(z)$, Eq.(\r{gaugetr}), is
given by 
\bea
\l{gaugetr1}
\delta_X A_{\alpha}(z)&=&{E_{\alpha}}^M(\partial_M X+ig_{YM}[A_M,X]),\n\\
 &=&{E_{\alpha}}^M\nabla_M X,
\eea
where the scalar superfield $X(z)$ is a gauge parameter of $U(N)$:
\[ X(z)=\omega(\sigma)+i{\bar \theta}\zeta(\sigma)+
\frac{i}{2}{\bar \theta}\theta \rho(\sigma).\] 
As shown in the Appendix B, the Wess-Zumino (WZ) gauge \c{wb},
$\eta=n=0$, can be chosen by using the gauge freedoms $\zeta$ and
$\rho$ and in this gauge the transformations (\r{gaugetr1}) of 
the component fields $v_m,\,\phi$, and $\lambda$ reduce to the ordinary
gauge transformations, so that they correspond to the Yang-Mills gauge
field, adjoint scalar, and their superpartner, gaugino, respectively.

Next we consider the superspace transformation of the superfields 
$A_{\alpha}(z)$ defined as \c{wb}
\be
\l{strvec}
\delta A_{\alpha}(z)=\xi^M\partial_M A_{\alpha}(z)
-\frac{1}{2}(\gamma_5)_{\alpha}\,^{\beta}A_{\beta}(z)L(z).
\ee
Since the above superspace transformation does not preserve the WZ
gauge $\eta=n=0$, the gauge parameter $X(z)$ should be field-dependent
and decomposed as
the following form in order to preserve the WZ gauge  
\be
\l{gaugepara}
X(z)=\omega(\sigma)+i{\bar \alpha}\gamma^m\theta v_m+
i{\bar \alpha}\gamma_5\theta \phi+\frac{1}{2}{\bar\theta}\theta
({\bar \alpha}\lambda-\frac{1}{2}{\bar \alpha}\gamma^m\gamma^n\chi_m v_n
+\frac{1}{4}{\bar \alpha}\gamma_5\gamma^m\chi_m\phi).
\ee
In the WZ gauge the superfield $A_{\alpha}(z)$ has the following 
component expression:
\be
\l{wzvm}
A_{\alpha}(z)=i(\gamma^m\theta)_{\alpha} v_m(\sigma)
+i(\gamma_5\theta)_{\alpha} \phi(\sigma)
+{\bar \theta}\theta 
(\lambda-\frac{1}{4}\gamma^m\gamma^n\chi_m v_n)_{\alpha}(\sigma).
\ee 
The supersymmetry transformation (\r{strvec}) for the Yang-Mills multiplet 
$(v_m,\,\phi,\,\lambda)$ becomes covariantized transformation in the
WZ gauge and can be determined as (see Appendix B)
\bea
\l{stym}
&&\delta v_m=i{\bar \alpha}\gamma^m\lambda
-\frac{i}{2}{\bar \alpha}\gamma_5\chi_m \phi
-\frac{i}{4}{\bar \alpha}\gamma_5\gamma^n\gamma_m\chi_n \phi,\n\\
&&\delta \phi=i{\bar \alpha}\gamma_5\lambda
-\frac{i}{4}{\bar \alpha}\gamma^m\chi_m \phi,\n\\
&&\delta \lambda=\frac{1}{2}\gamma^m\gamma^n\alpha (v_{nm}
+\frac{i}{2}{\bar \lambda}\gamma_m\chi_n
-\frac{i}{2}{\bar \lambda}\gamma_n\chi_m)
+\gamma_5\gamma^m\alpha (\nabla_m \phi 
+\frac{i}{2}{\bar \lambda}\gamma_5\chi_m)\\
&&\;\;\;\;\;\;\;\;+\frac{1}{2}\gamma_5\gamma^m({\cal D}_m\alpha 
-\frac{1}{4}\gamma_m\alpha A)\phi 
+\frac{i}{4}{\bar \alpha}\gamma_5\lambda\;
\gamma_5\gamma^m\chi_m \n\\
&&\;\;\;\;\;\;\;\;+\frac{i}{16}{\bar \chi}_n\gamma_5 
\gamma^n\gamma^m\gamma^l\chi_m \;\gamma_l\alpha\phi\;
-\frac{i}{32}{\bar \chi}_n\gamma^n\gamma^m\chi_m
\;\gamma_5\alpha\phi,\n
\eea
where $v_{nm}=\partial_n v_m-\partial_m v_n+ig_{YM}[v_n,\,v_m]$ is the
Yang-Mills field strength.

\section{Non-Abelian Ramond-Neveu-Schwarz String Theory}

According to the motivations mentioned in the Introduction, 
we will now try to construct consistent worldsheet formulation of 
RNS string theory where the ``spacetime coordinates''
of strings are treated as non-commuting matrices. Our string
coordinates $X^I(\sigma)$ are $N\times N$ hermitian matrices 
in the adjoint of an $U(N)$ group, carring also an $SO(D-1,1)$ vector 
index $I=0,1,\cdots,D-1$.\footnote{In next section we will give the 
reliable evidences that the critical dimension of the NARNS string 
is also 10 in a special limit.} 

In order to obtain a desirable formulation 
for the NARNS string, the adjoint matters in the gauge group $U(N)$, 
playing a role of matrix string coordinates, need to be introduced. 
Of course, this adjoint matter couples to the SYM field as well as
the worldsheet supergraviton. We thus introduce worldsheet scalar 
superfields $V^I(z)$ in the adjoint representation of the $U(N)$
group, which have the expansion:
\be
\l{VX} 
V^I(z)=X^I(\sigma)+i{\bar \theta}\psi^I(\sigma)+
\frac{i}{2}{\bar \theta}\theta F^I(\sigma).
\ee
Note that the worldsheet spinor $\psi^I(\sigma)$ is an $SO(D-1,1)$ 
vector like as $X^I(\sigma)$.
Under the superspace transformation, $V^I(z)$  changes as follows:
\[ \delta V^I(z)=\xi^M \partial_M V^I+ig_{YM}[V^I, X]\]
with the gauge parameter $X(z)$ given by Eq.(\r{gaugepara}).
Inserting the explicit form (\r{tpxi}) for $\xi^M$, 
one finds the supersymmetry transformation law for the component fields:
\bea
\l{stov}
&&\delta X^I=i{\bar \alpha}\psi^I\cr
&&\delta \psi^I=\gamma^m\alpha (\nabla_m X^I
-\frac{i}{2}{\bar \chi}_m\psi^I)+\alpha F^I +ig_{YM}\gamma_5\alpha[\phi, X^I]\\
&&\delta F^I=i{\bar \alpha}\gamma^m \Delta_m\psi^I
-\frac{i}{2}{\bar \alpha}\gamma^n\gamma^m\chi_n
(\nabla_m X^I-\frac{i}{2}{\bar \chi}_m\psi^I)
-\frac{i}{2}{\bar \alpha}\gamma^m\chi_m F^I\n\\
&&\;\;\;\;\;\;\;\;\;-g_{YM}{\bar \alpha}[\lambda, X^I]
-g_{YM}{\bar \alpha}\gamma_5[\phi, \psi^I]
-\frac{1}{4}g_{YM}{\bar \alpha}\gamma_5\gamma^m\chi_m[\phi, X^I].\n
\eea

We endow our system with the natural metric which is invariant 
under $U(N)\times SO(D-1,1)$:
\be
\l{metric}
|\delta X|^2=\int d^2\sigma g_{IJ}Tr(\delta X^I \delta X^J)
\ee
where $g_{IJ}$ is the Minkowski metric of the embedding space
with signature (D-1,1). 

Under the metric (\r{metric}), the gauge invariant locally 
supersymmetric worldsheet action is then given by
\be
\l{sac}
I_{WS}=\frac{T}{2}\int d^2zE g_{IJ}
Tr(E_\alpha\,^M\nabla_M V^I E^{\alpha N}
\nabla_N V^J),
\ee
where $d^2 z=d^2\sigma d\theta d{\bar \theta}$ is the superworldsheet
volume and $\nabla_M V^I=\partial_M V^I + ig_{YM}[A_M, V^I]$. 
The coupling constant $g_{YM}$ should be proportional to 
the string tension as 
$\sqrt T =1/\sqrt{2\pi \alpha'}$ for a dimensional
reason. Thus we introduce the dimensionless coupling constant $g_s$ as
\[ g^{-2}_{YM}\equiv g^2= \alpha' g_s^2.\]
Then the weak coupling is characterized by the condition 
$g_s^2= 2\pi T g_{YM}^{-2} \ll 1$. 

In Sec.II, we observed that the spinor superfields $A_\alpha(z)$ 
may be viewed as the supersymmetric generalization 
of the Yang-Mills potentials. 
To construct the corresponding SYM theory,  
we need a gauge covariant supersymmetric field strength related with 
the usual Yang-Mills part. It is easily confirmed that 
this object is given by the covariant derivative of 
the superfield $W$ in the Eq.(\r{a6}). 
Hence we can easily write down the SYM action 
coupled to 2D supergravity:
\be
\l{sym}
I_{YM}=-\frac{1}{8}\int d^2zE Tr(E_\alpha\,^M\nabla_M W E^{\alpha N}
\nabla_N W).
\ee 

We must emphasize the fact that the ``super Euler number'' 
defined on a super Riemann surface 
${\cal M}$ reduces to the standard Euler number
\be
\l{euler}
\chi({\cal M})=\frac{i}{2\pi}\int_{{\cal M}}d^2z ES=
\frac{1}{2\pi}\int_{{\cal M}}d^2\sigma eR,
\ee
where $R=\frac{\epsilon^{mn}}{e}\partial_m \omega_n$ is the curvature
of the connection $\omega_m$ in Eq.(\r{curv}). 
This action contains no term involving the auxiliary field $A$ 
unlike the higher dimensional supergravity \c{wb}. 
We will not consider 
the 2D pure supergravity action (\r{euler}) 
since it is a total derivative
and so have no dynamics. 

From the Eq.(\r{sac}) and Eq.(\r{sym}), one can obtain the 
corresponding worldsheet action for the NARNS strings
\bea
\l{nansr}
I= \int d^2\sigma e &&Tr\{\frac{T}{2}g_{IJ}(-g^{mn}
\nabla_m X^I \nabla_n X^J
-i{\bar \psi}^I \gamma^m \nabla_m \psi^J 
+i{\bar \chi}_n\gamma^m \gamma^n \psi^I\nabla_m X^J \n\\ 
&&-\frac{1}{8}{\bar \chi}_m\gamma^n\gamma^m \chi_n{\bar \psi}^I \psi^J
+ F^IF^J +\frac{1}{2g^2}[\phi,X^I][\phi,X^J]
+\frac{1}{g}{\bar \psi}^I[\lambda,X^J] \n\\
&&+\frac{1}{g}{\bar \psi}^I\gamma_5[\phi,\psi^J]
+\frac{1}{4g}{\bar \psi}^I\gamma_5\gamma^m\chi_m\;[\phi, X^J])
-\frac{g^2}{4}v_{mn}v^{mn}
-\frac{i}{2}{\bar \lambda}\gamma^m \Delta_m\lambda \n\\
&&-\frac{1}{2}g^{mn}\nabla_m\phi\nabla_n\phi
+\frac{i}{4}g{\bar \lambda}\gamma_5\gamma^m\chi_m\; \upsilon 
-\frac{i}{4}{\bar \lambda}\gamma_5\chi^m\; \nabla_m\phi
-\frac{3i}{8}{\bar \lambda}\gamma_5\gamma^n\gamma^m\chi_n\;\nabla_m\phi \n \\
&&-\frac{i}{8}{\bar \lambda}\gamma_5\gamma^m\gamma^n
{\cal D}_m\chi_n\;\phi
-\frac{i}{8}{\bar \chi}_n\gamma^n\gamma_5\gamma^m\Delta_m\lambda \;\phi
-\frac{i}{8}{\bar \chi}_p \gamma^p \chi^m\; \phi \nabla_m \phi
+\frac{g}{4} A \upsilon \phi \n\\
&&+\frac{i}{32}{\bar \chi}_p\gamma^p\gamma^m\gamma^n
{\cal D}_m \chi_n\; \phi \phi
+\frac{3}{32}{\bar \chi}_m\gamma^n\gamma^m\chi_n\;
{\bar \lambda}\lambda -\frac{i}{16}{\bar \chi}_n \gamma^m \gamma^n
\chi_m\; A \phi\phi \n \\
&&+\frac{3i}{16}{\bar \lambda}\gamma_5\gamma^m \chi_m\; A \phi
+\frac{i}{32}g{\bar \chi}_n\gamma^n \gamma^m \chi_m\; \upsilon \phi
+\frac{1}{64}{\bar \chi}_m\gamma^n\gamma^m\chi_n
\;{\bar \chi}^p\chi_p\; \phi\phi \n\\
&&+\frac{3}{64}{\bar \chi}_m\gamma^n\gamma^m\chi_n
\;{\bar \lambda}\gamma_5\gamma^p\chi_p\;\phi
+\frac{1}{8}A^2 \phi\phi+\frac{1}{g}{\bar
  \lambda}\gamma_5\lambda\;\phi\},
\eea
where $\upsilon=\frac{1}{e}\epsilon^{mn}v_{mn}$ and 
${\cal D}_m\chi_n \equiv e_{na}{\cal D}_m\chi^a=e_{na}(\partial_m \chi^a-
\frac{1}{2}\omega_m\gamma_5\chi^a+\omega_m\epsilon^{ab}\chi_b)$. 
The action (\r{nansr}) is automatically invariant under supersymmetry 
transformations, Eq.(\r{stgm}), Eq.(\r{stym}), and Eq.(\r{stov}) 
(up to total derivatives) 
because it was derived from a superspace formalism.
The Yukawa type interaction, $g_{YM}{\bar
\lambda}\gamma_5\lambda\;\phi$, 
in Eq.(\r{nansr}) is the only term coming 
from the commutator in Eq.(\r{sym}).
The Yang-Mills coupling $g_{YM}$ has been absorbed in the
normalization of $v_m: v_m \rightarrow \frac{v_m}{g_{YM}}$.
  
In two dimensions, the Yang-Mills gauge fields themselves have 
no propagating degrees of freedom - there are no gluons. 
This does not make the theory trivial, but the gauge field interactions
give rise to a confining potential for colored objects \c{'t Hooft}. 
In addition, the SYM multiplet in two dimensions contains genuine 
dynamical degrees of freedom in the adjoint representation. 
Thus 2D SYM theory may reveal nontrivial physical 
spectrums such as nonperturbative vacuum structures \c{sakai}. 
Although all this is true, the SYM action (\r{sym}) is at most order of
$\alpha'$ compared to the worldsheet action (\r{sac}). 
In infrared (IR) limit, i.e. $\alpha'\rightarrow 0$ with fixed $g_s$, 
the SYM part can be thus ignored. 
It is important to observe that, in this limit, 
the SYM action (\r{sym}) is strongly coupled and we expect a
nontrivial CFT to describe the IR fixed point.

It turns out that we can find this CFT description via the following 
rather naive reasoning. We first notice that, in the 
$\alpha'\rightarrow 0$ limit, the potential terms comming from the
commutators in the NARNS string acton (\r{nansr}) effectively turn
into constraints, requiring all the matrix fields in the adjoint
representation of $U(N)$ to commute.\footnote{Although the constraint
$[\phi,X^I]^2=0$ does not necessarily imply $[\phi,X^I]=0$ because of the 
$SO(D-1,1)$ metric $g_{IJ}$, we will ignore the possibility.}
This means that we can write the matrix coordinates $X^I$ in a
simultaneously diagonalized form 
\be
\l{xc}
X^I=U^{-1} \mbox{diag}(x^I_1, \cdots, x^I_N)U
\ee
with $U \in U(N)$. Here $U$ and all eigenvalues $x^I_a$ can of course
still depend on the worldsheet coordinates. 
In this IR limit our NARNS string theory has a free
string limit where the usual RNS string theory is recovered and 
it becomes N-copies of usual RNS string \c{bdhz}:
\bea
\l{nsr}
I_{RNS}=\frac{T}{2} \int d^2\sigma &&e g^{IJ}(-g^{mn}
\partial_m x_I^a \partial_n x_J^a
-i{\bar \psi}_I^a \gamma^m \partial_m \psi_J^a 
+i{\bar \chi}_n\gamma^m \gamma^n \psi_I^a\partial_m x_J^a \n\\
&&-\frac{1}{8}{\bar \chi}_m\gamma^n\gamma^m \chi_n{\bar \psi}_I^a \psi_J^a
+ F_I^aF_J^a),
\eea 
where $a=1,\cdots, N$ is a Lie algebra index belonging only to Cartan
subalgebra of $U(N)$. This action is invariant under general
coordinate transformations, local Lorentz transformation, local
supersymmetry transformations, Eq.(\r{stgm}) and Eq.(\r{stov}) 
(with $g_{YM}=0$), 
and Weyl transformation, under which the fields rescale as follows 
($\Lambda=\Lambda(\sigma)$)
\be
\l{weyl}
x_I^a \rightarrow x_I^a, \;\;\;\; \psi_I^a \rightarrow  
\Lambda^{-1/2} \psi_I^a, \;\;\;\; F_I^a \rightarrow  
\Lambda^{-1} F_I^a, \;\;\;\; 
{e_m}^b \rightarrow \Lambda {e_m}^b, \;\;\;\;
\chi_m \rightarrow  \Lambda^{1/2} \chi_m.
\ee
Furthermore the action is invariant under the following transformation
\be
\l{ss}
\chi_m \rightarrow \chi_m +\gamma_m \chi,
\ee
where $\chi(\sigma)$ is an arbitrary Majorana spinor. 
Using these eight gauge degrees of freedom, ${e_m}^a$ and
$\chi_m$ can be completely gauged away. 
In other words, we may choose the superconformal gauge, 
${e_m}^a={\delta_m}^a$ and
$\chi_m=0$ \c{bdhz}. 
Through the equation of motion of the auxiliary 
field $F^a_I$, we then arrive at the gauge fixed Polyakov action 
\c{polyakov} 
whose quantum theory is given by superconformal field theory (SCFT)
\be
\l{pa}
I_P=\frac{T}{2} \int d^2\sigma g^{IJ}(-\eta^{mn}
\partial_m x_I^a \partial_n x_J^a
-i{\bar \psi}_I^a \gamma^m \partial_m \psi_J^a),
\ee
where $\eta^{mn}$ is a flat worldsheet metric.
  
It is interesting to observe that, in the $g_s \rightarrow 0$ limit 
with fixed $\alpha'$, 
the SYM action (\r{sym}) is invariant under a Weyl transformation that
rescales the fields according to
\be
\l{ymweyl}
\phi \rightarrow \phi, \;\;\;\; \lambda \rightarrow  
\Lambda^{-1/2}\lambda, \;\;\;\; v_m \rightarrow v_m
\ee 
together with the super Weyl transformation of $E_M\,^A$ which
consists of the last two equations in Eq.(\r{weyl}) and 
$A \rightarrow \Lambda^{-1} A$ \c{howe}. 
This fact can be most easily understood as following way: 
Since the terms involving with $v_{mn}$ vanish in the
limit, the superfield $W$ in 
Eq.(\r{sym}) exhibits the same behavior as the 
superfield $V^I$ in Eq.(\r{sac}) 
under the conformal transformation (\r{ymweyl}). In addition, the Yukawa
type interaction which breaks the conformal invariance vanishes 
in the limit because the gaugino $\lambda$ and the scalar $\phi$ should
align in the direction of Cartan subalgebra of $U(N)$. 
Then the behavior of the SYM action under
the super Weyl transformation is exactly the same as the action 
(\r{sac}), which is superconformally invariant as proved by Howe \c{howe}. 
This implies that the Weyl symmetry is preserved at least up to
$\alpha'$ order provided that the SYM multiplet scales as 
the Eq.(\r{ymweyl}).   
But the transformation (\r{ss}) is no more symmetry of the action
(\r{sym}). It will be showed that this symmetry is recovered at a 
particular situation.

When the SYM action in the $g_s \rightarrow 0$ limit is considered, 
we can use the only six gauge degrees 
of freedom \footnote{Although the NARNS string theory definitely 
has an additive $U(N)$ gauge symmetry, we would, for the present, 
keep the $U(N)$ symmetry.} 
and so may choose the gauge ${e_m}^a={\delta_m}^a$ and
$\chi_m=\gamma_m \chi$. In this gauge, the NARNS string action
(\r{nansr}) becomes
\bea
\l{weak}
I_{Weak}=T\int d^2\sigma &&Tr\{\frac{1}{2} g^{IJ}(-\eta^{mn}
\nabla_m X_I \nabla_n X_J
-i{\bar \psi}_I \gamma^m \nabla_m \psi_J)\n\\
&&+2\pi\alpha' (-\frac{1}{2}\eta^{mn}\nabla_m\phi\nabla_n\phi
-\frac{i}{2}{\bar \lambda}\gamma^m \nabla_m\lambda \\
&&-\frac{i}{4}{\bar \chi}\gamma^m
\partial_m \chi\; \phi \phi
+\frac{3i}{8}{\bar \lambda}\gamma_5\chi\; A \phi
+\frac{1}{8}A^2 \phi\phi)\}.\n
\eea
Note that, in the limit $g_s \rightarrow 0$, all the
matrix fields in the action (\r{weak}) 
except the Yang-Mills gauge fields $v_m$ still take the diagonalized
form such as $X^I$ in the Eq.(\r{xc}). In order to remove the 
auxiliary field $A$, we use the equation of motion for $A$
\[A= -\frac{3i}{2}{\bar \lambda}\gamma_5\chi \; \phi / \phi\phi.\]
Then the action $I_{Weak}$ reduces to the following form
\bea
\l{rweak}
I_{Weak}=&&T \int d^2\sigma Tr\{\frac{1}{2}g^{IJ}(-\eta^{mn}
\nabla_m X_I \nabla_n X_J
-i{\bar \psi}_I \gamma^m \nabla_m \psi_J) \n\\
&&\;\;\;\;\;\;\;\;\;\;\;\;\;\;\;\;\;
+2\pi\alpha' (-\frac{1}{2}\eta^{mn}\nabla_m\phi\nabla_n\phi
-\frac{i}{2}{\bar \lambda}\gamma^m \nabla_m\lambda)\}\\
&&+\frac{T}{4}\int d^2\sigma 2\pi\alpha'Tr(-i{\bar \chi}\gamma^m
\partial_m \chi\; \phi \phi
+\frac{9}{16}{\bar \chi}\chi\;{\bar{\tilde \lambda}}{\tilde \lambda}),\n
\eea
where ${\tilde \lambda} =\lambda^a {\hat \phi}^a,\;\; 
{\hat \phi}^a=\phi^a/\sqrt{\phi\phi}$. As usual, we have the
constraints comming from the equations of motion for the zweibein and
gravitino due to the above gauge fixing:
\bea 
\l{tf}
T_{mn}&=&\frac{1}{2}Tr\{\nabla_m X^I \nabla_n X_I+
i{\bar \psi}^I \gamma_{(m} \nabla_{n)} \psi_I
+2\pi\alpha'(\nabla_m\phi\nabla_n \phi
+i{\bar \lambda}\gamma_{(m} \nabla_{n)}\lambda 
+\frac{i}{8}{\bar \chi}\gamma_{(m}\partial_{n)} \chi\; \phi \phi)\} \n\\
&-&\frac{1}{4}\eta_{mn} Tr\{\nabla^p X^I \nabla_p X_I
+i{\bar \psi}^I \gamma^p \nabla_p \psi_I
+2\pi\alpha' (\nabla^p\phi\nabla_p\phi
+i{\bar \lambda}\gamma^p \nabla_p\lambda)\},\\
F_m &=& \frac{i}{2}Tr(\gamma^n \gamma_m \psi^I\;\nabla_n X_I)
+\frac{1}{4}\pi \alpha'Tr (3i\gamma_5 \gamma^n \gamma_m \lambda 
\;\nabla_n \phi -2i \partial_m \phi^2 +i\gamma_m \gamma^n \partial_n\chi
\;\phi^2 +\frac{9}{8}{\bar{\tilde \lambda}}{\tilde \lambda} \;\chi).\n
\eea   
The superconformal generators, $T_{mn}$ and $F_{m}$, no loger satisfy
the tracelessness property
\bea
\l{trace}
&&T_{m}^{m}=\frac{i}{16}{\bar \chi}\gamma^m\partial_m \chi\; \phi
\phi,\n\\
&&\gamma^m F_{m}=2\pi \alpha' (\frac{i}{4} \gamma^m\chi 
\; \partial_m \phi^2 +\frac{i}{4}\gamma^m \partial_m\chi
\;\phi^2 +\frac{9}{64}\gamma^m\chi \;{\bar{\tilde \lambda}}{\tilde \lambda}).
\eea  
This is due to the fact the NARNS string action in the $g_s
\rightarrow 0$ limit does not have the symmetry (\r{ss}), thus full
superconformal symmetry.  

We would like to seek the particular situation for the superconformal
symmetry of the NARNS string to be recovered. In order to satisfy this
requirement, we must have the conditions, $T_m^m=\gamma^m F_m=0$. 
Interestingly, this can be achieved if only the following conditions
are fufilled
\be
\l{sucon}
{\tilde \lambda} =\lambda^a {\hat \phi}^a=0, \;\;\;
\phi^2= \phi^a \phi^a=\mbox{constant},
\ee
and together with the Dirac equation $\gamma^m\partial_m\chi=0$.
In order for the constraint $\lambda^a {\hat \phi}^a=0$ 
not to completely break the worldsheet supersymmtry (\r{stym}), 
we should have an another condition
\[{\bar \lambda}^a\lambda^a+\frac{1}{4} {\bar\chi}\chi 
\; \phi^a\phi^a=0.\]
Together with the condition (\r{sucon}), this implies that the gaugino
and the scalar field must parallely align in the $U(N)$ group space:
\be
\l{align}
\lambda^a=\pm \frac{1}{2} \chi\; \phi^a.
\ee
When the condition (\r{align}) is satisfied, we arrive at a 
configuration to recover the superconformal symmetry
\be
\l{11}
\lambda^a=\chi=0,\;\;\;\phi^2= \phi^a\phi^a=\mbox{constant}.
\ee
Then the scalar field $\phi$ in the Yang-Mills multiplet becomes a
singlet with respect to the worldsheet supersymmetry, 
i.e. $\delta \phi=0$, and behaves as
a modular parameter of the theory. 

Let us give some remarks for this interesting phenomenon. First,
notice that the SYM gauge theory in two dimensions 
can be obtained by a dimensional
reduction from the SYM gauge theory in {\it three} dimensions. 
The adjoint scalar field can be understood as the component of the
gauge field in the compactified dimension. Note that the Yukawa 
interaction in the Eq.(\r{nansr}) is nothing but the gauge 
interaction in this compactified extra dimension. 
Second, from the Eq.(\r{rweak}), we observe that the scalar field
$\phi$ in the Yang-Mills multiplet shows the same behavior as the
string coordinates $X^I$, i.e., it behaves as if it is an another
coordinate of the string. If we would interpret the scalar field as the
field living on a some compactified dimension, we can introduce a  
new coordinate along this compactified direction as 
\be
\l{tdual}
X_{D+1}=\sqrt{2\pi\alpha'}\;\phi.
\ee
If the Abelianized string theory (\r{nsr}) is defined in the ten
dimensions, the coordinate $X_{D+1}$ in the Eq.(\r{tdual}) should be
involved with the {\it eleven} dimensional coordinate. 
Then the condition (\r{11}) constrains that the eleven dimensional 
coordinate $X_{D+1}$ should be defined on an (N-1)-dimensional 
orbifold $S^{N-1}/S_N$, where the orbifold group $S_N$ comes from the
discrete Weyl symmetry of $U(N)$. 
It is prudently expected that our NARNS string theory has the natural 
M-theory interpretation and is related with the Matrix string theory.

\section{NARNS String As Matrix String Theory}
In the previous section we showed that our NARNS string theory 
in the $\alpha'\rightarrow 0$ limit with fixed $g_s$ has a free
string limit where the usual RNS string theory is recovered and 
it becomes N-copies of usual RNS string. And we observed that, 
in the weak coupling limit, i.e. $g_s \rightarrow 0$, 
a new additional dimension
appears in the string spectrum and it can be speculatively interpreted
as the compactified {\it eleven} dimensional coordinate. 
In this section, we will argue that the NARNS string theory 
in the $\alpha'\rightarrow 0$ limit with fixed $g_s$ 
can be described by the orbifold conformal 
field theory, which seems to correspond to the manifestly covariant 
worldsheet version of the MST of 
Dijkgraaf, Verlinde and Verlinde (DVV) \c{DVV}. And we observe that, 
in the weak coupling limit, i.e. $g_s \rightarrow 0$, 
the dynamics of a new additional {\it eleven} dimensional coordinate 
is given by an $S_N$-orbifold O(N) sigma model. Our speculation in this
section are preliminary and conjectural. We hope our conjectural
speculations to be completed by the detailed analysis in the near future.  

Let us recall the M(atrix) formulation of M-theory by 
BFSS \c{BFSS} and Matrix string theory by DVV \c{DVV}. Consider
an M-theory excitation on ${\bf R}^{10}\times S^1$ 
with finite mass $m$ which satisfies a Lorentz
invariant eleven dimensional dispersion relation
\be
\l{11d}
 -P^M P_M=E^2-P_{11}^2-{\vec P}_{\bot}^2=m^2,
\ee
where the eleven dimensional momentum $P_{11}$ along the circle $S^1$
with radius $R_{11}$ should be quantized as $P_{11}=N/R_{11}$. 
In an IMF boosted along the $S^1$, 
$P_{11}\rightarrow \infty$, and a
particular decompactified limit, i.e. $R_{11}\rightarrow \infty$, 
the dispersion relation (\r{11d}) with the eleven dimensional 
Lorentz invariance effectively reduces to the transverse 9-dimensional
Galilean dynamics:
\be
\l{d0}
H_{\bot}=E-P_{11}=\frac{{\vec P}_{\bot}^2}{2P_{11}}.
\ee    
This relation implies that M-theory quantum dynamics in IMF may be
captured by the quantum mechanics of particles with the mass
$m=P_{11}$. What is the M-theory object on ${\bf R}^{10}\times S^1$ 
carrying the mass $P_{11}$? 
According to Witten \c{Witt95}, 
we see that this is just a D0-brane, Kaluza-Klein
excitation along the eleven dimensional circle $S^1$, whose IR dynamics
is given by the super-Yang-Mills quantum mechanics \c{D0QM} 
defined in 9 space 
dimensions ${\bf R}^{9}\;(X^i,\; i=1,\cdots,9)$. 
Thus the following conjecture can be made:
\[M\mbox{-}theory\; quantum\; dynamics\; in\; IMF\; is\; described \;by\;
the\; D0-brane\;quantum\; mechanics.\] 
This is exactly the BFSS conjecture \c{BFSS} which almost has passed many
nontrivial tests so far. 

What Witten has shown is that the strong coupling limit of type 
IIA superstring theory should be identical to the M-theory 
on ${\bf R}^{10}\times S^1$ \c{Witt95}. According to this ficture, 
if we consider the BFSS matrix theory more compactified on a circle 
$S^1$ of radius $R_{9}$ along the $X^9$ direction, and if we think of
dimension 9 rather than dimension 11 as the M-theory compactification
direction to get the type IIA theory, the SYM theory in
two dimensions should provide a light-front description of 
the type IIA string theory according to the duality relation \c{taylor}
\be
\l{dual}
IIA\;\; \sim \;M\; on\; S^1\sim \;\;M(atrix)\;on\; {\tilde S}^1,
\ee
where the radius ${\tilde R_9}$ of the dual circle ${\tilde S}^1$ is
related by ${\tilde R_9}=1/2\pi R_9$. 
Since the dimension 9 is the M-theory compactification
direction, the fumdamental objects which carry the light-front
momentum $p^+=N/R_{11}$ are no longer D0-branes, but rather strings. 
Thus this gives the DVV description on MST \c{DVV}, namely
that 2D Yang-Mills theory in the large N limit should
correspond to light-front type IIA string theory which is given by a
sigma model on the orbifold target space $({\bf R}^{8})^N /S_N$ as 
$R_9 \rightarrow 0$. The Weyl symmetry $S_N$ is the discrete remnant 
of the gauge group $U(N)$ acting within the Cartan subalgebra, 
indicating that the string bits, partons carrying a minimum unit 
of light-front momemtum, should
be treated as indistinguishable objects. 
One particularly nice
picture of MST comes in following way. Since the
string configuration may respect the residual gauge symmetry $S_N$
if we go around the space-like $S^1$ of the worldsheet, 
the matrix configuration 
need not be periodic in $\sigma$. The matrices $X^i(0)$ and
$X^i(2\pi)$ can be related by an arbitrary permutation. The lengths of
the cycles of this permutation determine the numbers of string bits, 
which combine into long strings whose longitudinal momentum 
$p^+=n/R_{11}$ can become large in the large N limit. The twisted
sectors of this theory correspond precisely to the sectors where the
string bits are combined in different permutations.
      
We want to construct
a theory with ``space-time'' Poincar\'e symmetry as well as $U(N)$
gauge symmetry. The Poincar\'e symmetry for the matrix coordinates $\{X_I\}$
means the global $SO(D-1,1)$ symmetry plus the translations acting on
the matrices $X_I$ by $X_I \rightarrow X_I + a_I \cdot I_{N \times N}$. 
The NARNS string action (\r{nansr}) manifestly has this 
global ``space-time'' Poincar\'e  symmetry. 
The motion of the center of mass of the system associated
with the global shift $X_I \rightarrow X_I + Tr{X_I}/N \cdot I_{N \times N}$ 
comes from the $U(1)$ part by separating off the trace
part of $U(N)$, i.e. $U(N)=U(1)\times SU(N)$. Thus $SU(N)$ matrices
describe the relative motion of the system \c{BFSS}.

In this section we take string unit $\alpha'$=2 and the two
dimensional worldsheet is taken to be a cylinder parameterized by the
coordinates $(\tau, \sigma)$ with $\sigma$ between $0$ and 2$\pi$. 
The light-cone matrix coordinates are defined to be 
$X^{\pm}=\frac{1}{\sqrt{2}}(X^0\pm X^{D-1}),\;
\psi^{\pm}=\frac{1}{\sqrt{2}}(\psi^0\pm \psi^{D-1})$ and
$X^i,\,\psi^i,\;i=1,\cdots,D-2$ and the scalar product 
in terms of light-cone 
components is $V\cdot W = -V^+W^- -V^-W^+ +V^iW^i$.

In IR limit, i.e. $\alpha'\rightarrow 0$ with fixed $g_s$,\footnote{We
have taken the string unit $\alpha'=2$. Here the $\alpha' \rightarrow
0$ limit means that the typical length scale under consideration is very
very large compared to the string scale $\alpha'=2$.} 
where $g_{YM} \rightarrow \infty$, since 
the 2D Yang-Mills theory is also a confining phase 
\c{'t Hooft}, only
observable entries, that can escape the confining
potential, are diagonalized matrix fields. 
Thus the usual spacetime picture emerges when 
all of the $X_I$ commute with each
other, hence can be simultaneously diagonalized as the
indistinguishable $N-$tuple points $X^I=\mbox{diag}(x^I_a)$.
In this limit, the SYM
part disappears in the string spectrum and the $U(N)$ gauge symmetry 
is generically broken down to $U(1)^N$ with the Weyl group, 
the residual discrete symmetry of the
gauge group, acting on the eigenvalues in Eq.(\r{xc}).
Therefore the light-cone SCFT of 
the Polyakov action (\r{pa}) is given by the N-tuple transverse fields
$x_i^a,\;\psi_i^a,\;i=1,\cdots,D-2,\;a=1,\cdots,N$ which define the
orbifold target space given by the symmetric product space
\be
\l{orbifold}
S^N{\bf R}^{D-2}= ({\bf R}^{D-2})^N/S_N.
\ee
Since the NARNS string theory in the IR limit is also described by the
$S_N$-orbifold CFT, we can follow the exactly same route taken 
by \c{DMVV} and \c{DVV}, keeping in mind some issues such as the GSO
projection \c{GSO} and the level-matching condition \c{string} 
discussed later. 
In the Ref.\c{DMVV}, it is shown that this orbifold conformal field
theory exactly corresponds to a second quantized string theory 
on the space ${\bf R}^{D-2} \times S^1$. Based on the exact
equivalence between a second-quantized string spectrum and the
spectrum of a 2D $S_N$-orbifold sigma model, this
correspondence is more elaborated in \c{DVV} as Matrix string theory 
described above. 
In this correspondence, the Hilbert space of the orbifold conformal
field theory is decomposed into twisted sectors ${\cal H}_g$ labeled by
the conjugacy classes $[g]$ of the orbifold group $S_N$. The conjugacy
classes $[g]$ are characterized by partitions $\{N_n\}$ of N
\be
\l{partition}
\sum_n n N_n=N,
\ee
where $N_n$ denotes the multiplicity of the cyclic permutation $(n)$
of $n$ elements in the decomposition of $g$
\be
\l{congacy}
[g]= (1)^{N_1}(2)^{N_2}\cdots (s)^{N_s}.
\ee
In each twisted sector, one must further keep only the states
invariant under the centralizer subgroup $C_g$ of $g$, which takes the
form 
\be
\l{centralizer}
C_g=\prod_{n=1}^s S_{N_n} \times {\bf Z}_n^{N_n}.
\ee
Here each factor $S_{N_n}$ permutes the $N_n$ cycles $(n)$, while each
${\bf Z}_n$ acts within one particular cycle $(n)$. 
Thus the total orbifold Hilbert space takes the form 
\be
\l{orbhil}
{\cal H}(S^N{\bf R}^{D-2})=\bigoplus_{[g]} {\cal H}_g^{C_g},
\ee
where $C_g$ invariant subspace ${\cal H}_g^{C_g}$ can be decomposed
into the product over the subfactors $(n)$ of $N_n$-fold symmetric
tensor products of appropriate smaller Hilbert spaces 
${\cal H}_{(n)}^{{\bf Z}_n}$
\be
\l{znhilbert}
{\cal H}_g^{C_g}=\bigotimes_{n>0}S^{N_n} {\cal H}_{(n)}^{Z_n}.
\ee
The Hilbert spaces ${\cal H}_{(n)}^{{\bf Z}_n}$ in (\r{znhilbert}) denote
the ${\bf Z}_n$ invariant subspace of a single string on 
${\bf R}^{D-2}\times S^1$ with winding number $n$. 
We can represent this space using $n$ coordinate fields 
$x_a(\sigma) \in {\bf R}^{D-2}$ with the cyclic boundary condition
\be
\l{cyclicb}
x_a(\sigma +2\pi)=x_{a+1}(\sigma),
\ee
for $a \in (1, \cdots, n)$. We can glue the $n$ coordinate fields 
$x(\sigma)$ together into one single string field $x(\sigma)$ defined
on the interval $0\le \sigma \le 2\pi n$. Hence, the oscillators of
the long string that generate ${\cal H}_{(n)}^{{\bf Z}_n}$ have a fractional
$1/n$ moding relative to the string with winding number one. The group
${\bf Z}_n$ is generated by the cyclic permutation
\be
\l{cp}
\omega: x_a \rightarrow x_{a+1}
\ee
which via (\r{cyclicb}) corresponds to a translation $\sigma
\rightarrow \sigma+2\pi$. Thus the ${\bf Z}_n$-invariant subspace consists
of those states for which the fractional left-moving minus
right-moving oscillator numbers combined add up to an integer.

In the $S_N$-orbifold CFT, the twisted sectors of the orbifold
corresponding to the possible multistring states are all
superselection sectors unless string interactions are introduced that
generate the elementary joining and splitting of strings. 
Thus the GSO projection summing over all possible spin structures for
the string amplitudes \c{string} independently applies to each string in each
superselection sector. This projection is
performed separately on left and right movers because the states of
the closed NARNS superstring are direct products of the Fock space
states for the right and left movers. This degree of freedom
gives us two types of string theory, i.e. type IIA and type IIB string
theories. Then our NARNS string theory impartially should provide the matrix
formulation of both type IIA and type IIB string
theories. Since each string with lengths $n$ is the
same as the usual RNS strings with same lengths, this reasoning then leads
to the important conclusion that the critical dimension of 
the NARNS string theory in the IR limit is also 10 
in which case superconformal symmetry is
manifest even in the quantum level \c{polyakov}. 
So we will restrict to the case D=10.  

Let us compactify the light-like coordinate 
$X^{-}=\frac{1}{\sqrt{2}}(X^0- X^{D-1})$ on a circle of radius R,
which is essentially taken to infinity in order to obtain a
uncompactified limit. In this case the conjugate momentum $p^+$ is
quantized as $p^+=N'/R$ with $N'$ being integer valued.
This step introduces the twisted sectors corresponding to strings
wound around the peoriodically identified coordinate $X^-$, each
describing a set of noninteracting strings of length 
proportional to the carrying light-front momentum $n/R$ satisfying
$N_1+2N_2\cdots+n N_n=N'$. For an instructive discussion on discrete
light-cone quantization of string theory, see \c{Susskind}. 
Until now, it seems that there is no relation between the light-front
momentum $N'$ and the dimension of gauge group $N$. 
But, according to the string bit or parton picture, considering
the fact that the lengths $n$ of the individual strings specifies its
light-cone momentum and the string Fock space is characterized by an
integer $N'$ satisfying $\sum_n n N_n = N'$, it is resonable to
identify $N$ with $N'$. The invariance under ${\bf Z}_n$ implies that 
$N_L^{(n)}-N_R^{(n)}$ is a multiple of $n$ where 
$N_L^{(n)},\;N_R^{(n)}$ are the usual oscillator level numbers of the
strings with lengths $n$ \c{DVV}. In the limit $N \rightarrow \infty$, we
obtain the usual level matching conditions $N_L^{(n)}-N_R^{(n)}=0$
since all the string states for which $N_L^{(n)}-N_R^{(n)}\neq 0$
becomes infinitely massive at large N and $\alpha' \rightarrow 0$
limit. 
  
Now we will consider an another limit, $g_s \rightarrow 0$ limit
with fixed $\alpha'$, where the SYM part cannot be ignored. 
In the previous section, we showed that the configuration (\r{11}) 
preserves the full superconformal symmetry, so 
the superconformal gauge fixing can
be made: ${e_m}^a={\delta_m}^a$ and $\chi_m=0$. In addition we will
fix the $U(N)$ gauge symmetry as $v_m=0$.   
In this gauge the action (\r{rweak}) takes the CFT limit 
\be
\l{cft}
I_{CFT}=-\frac{1}{8\pi}\int d^2\sigma Tr \{\partial^m X^{I}\partial_m X_I
+i{\bar \psi}^{I} \gamma^m \partial_m \psi_I
+\frac{4\pi}{\lambda^2_\phi}\partial^m{\hat \phi}\partial_m{\hat \phi}\}
\ee 
with the generators
\bea 
\l{cftg}
T_{mn}&=&T_{mn}^{OCFT}+T_{mn}^{S^{N-1}}\n\\
&=&\frac{1}{2}Tr\{\partial_m X^{I} \partial_n X_I
+i{\bar \psi}^{I} \gamma_{(m} \partial_{n)} \psi_I
-\frac{1}{2}\eta_{mn} (\partial^p X^{I} \partial_p X_I
+i{\bar \psi}^{I} \gamma^p \partial_p \psi_I) \n\\
&&\;+\frac{4\pi}{\lambda^2_\phi}
(\partial_m{\hat \phi}\partial_n {\hat \phi}
-\eta_{mn}\frac{1}{2}\partial^p{\hat \phi}
\partial_p{\hat \phi})\},\\
F_m &=& \frac{i}{2}Tr(\gamma^n \gamma_m \psi^I\;\partial_n X_I), \n 
\eea 
where we have identified the coupling constant
$\lambda_\phi=1/\sqrt{\phi\phi}$ with the inverse radius 
of the sphere $S^{N-1}$. Note that the gauge currents 
$J_m^a \equiv \delta I_{Weak}/\delta v^m_a$ identically vanish because
the $U(N)$ guage symmetry was generically broken down 
to $U(1)^N$ in the limit under consideration. Note that we have still
discrete Weyl symmetry $S_N$ acting on the eigenvalues of matrix fields. 
After going to light-cone gauge $X^+ = 2 p^+ \tau$ (where $p^+$ is a
diagonalized N $\times$ N matrix),$\;\psi^+=0$ 
which fixes the gauge completely \c{string}, we can solve the constraints 
(\r{cftg}) to determine the coordinates $X^-$ and $\psi^-$ as
\bea
\l{lcc}
&&\delta_{m,0}\partial_n X^- +\delta_{n,0}\partial_m X^- 
+\eta_{nm}\partial_0 X^- \n\\
&&=\frac{1}{2}{p^+}^{-1}
Tr\{\partial_m X^i \partial_n X^i
+i{\bar \psi}^i \gamma_{(m} \partial_{n)} \psi^i
-\frac{1}{2}\eta_{mn} (\partial^p X^i \partial_p X^i
+i{\bar \psi}^i \gamma^p \partial_p \psi^i) \n\\
&&\;\;\;+\frac{4\pi}{\lambda^2_\phi}
(\partial_m{\hat \phi}\partial_n {\hat \phi}
-\eta_{mn}\frac{1}{2}\partial^p{\hat \phi}
\partial_p{\hat \phi})\},\\
&&\psi^- = \frac{1}{2}{p^+}^{-1}
Tr(\gamma^n \gamma^0 \psi^i\;\partial_n X^i),\n
\eea   
which leaves only the transverse components $X^i,\;\psi^i$ and ${\hat
\phi}$ as independent degrees of freedom (assuming that 
${\hat \phi} \in S^{N-1}/S_N$). In terms of these transversal 
degrees of freedom only, the light-cone action is simply
\be
\l{lc}
I_{LC}=-\frac{1}{8\pi}\int d^2\sigma Tr \{\partial^m X^i\partial_m X^i
+i{\bar \psi}^i \gamma^m \partial_m \psi^i
+\frac{4\pi}{\lambda^2_\phi}\partial^m{\hat \phi}\partial_m{\hat \phi}\}.
\ee 
The light-cone action (\r{lc}) consists of two parts. The N$\times$N
matrix fields $X^i$ and $\psi^i$ transform in the ${\bf 8}_v$ 
representation of the transversal SO(8) symmetry group. Apart from
these fields, we have the scalar field ${\hat \phi}$ defined 
on the orbifold $S^{N-1}/S_N$
which can be identified with the coset space $S^{N-1}/S_N \cong
SO(N)/SO(N-1) \times S_N$ as the target manifold. 
The diagonalized matrix fields $X^i$ and $\psi^i$ define 
the $S_N$-orbifold CFT on the N-fold symmetric product space 
$S^N{\bf R}^{8}= ({\bf R}^{8})^N/S_N$ which has been already discussed.    

The scalar fields ${\hat \phi}$ define so-called $S_N$-orbifold 
O(N) sigma model. The most simplest nontrivial example is a 
${\bf Z}_2$-orbifold $O(2)$ sigma model which has been extensively studied
so far \c{atmodel,dvv}. 
This model describes a free massless scalar field ${\hat
\phi}$ compactified on the orbifold $S^1/{\bf Z}_2$, known to describe the
critical line of the Ashkin-Teller model, i.e. two Ising models
coupled by a four spin interaction. 
Recall that the circle $S^1$ is just the quotient of an infinite 
line ${\bf R}$ by a discrete group $\Gamma = 2\pi R {\bf Z}$, i.e. 
$S^1 \cong {\bf R}/\Gamma$. This orbifold O(2) model exhibits a quite
interesting property, say that the $R=\sqrt 2$ torus model, 
compactified on a circle instead of an orbifold, which has an 
$SU(2) \times SU(2)$ symmetry and the $R=1$ orbifold model, which is
equivalent to two decoupled Ising models and consequently carries a
representation of two $c=\frac{1}{2}$ Virasoro algebras, is exactly
equivalent to each other \c{atmodel,dvv}. 
It is a kind of an electric-magnetic S-duality
relating the models at $R$ and $2/R$, while, in our case, 
it is a T-duality \c{tasi}. 

If a Wess-Zumino-Witten (WZW) term is included with the kinetic 
energy term, this theory can be described by the gauged 
WZW theory based on the coset model $SO(N)/SO(N-1) \times S_N$ using 
the Goddard, Kent, and Olive algebraic construction \c{GKO}. 
This model provides a interesting
property that the purely bosonic model is equivalent to a free fermion
theory at a particular infrared fixed point \c{Witt84}.
It has been known by the explicit analysis of O(4) model with no 
orbifold group \c{o4} that the
radius of $S^3$ increases with energy, and as a result the local
curvature tends to zero and in the infrared limit the radius decreases
with decreasing energy, but it stops decreasing as it reaches its
minimum critical radius which is the infrared fixed point having the
fermionic description. Although the presence of these interesting
phenomena, we don't completely identify the spectrum of this theory
with the string theory aspects, particularly involving with the
mysterious eleven dimensional M-theory.

\section{Conclusions}

According to the recent remarkable picture, so-called noncommutative 
spacetime geometry, appeared in nonperturbative string theory and
M(atrix)-theory, we, in this paper, considered NARNS superstring 
theory as a generalization of 
usual RNS string theory. It is a 2D supergravity 
theory coupled to SYM fields and adjoint matters
in a gauge group $U(N)$. 
Therefore the string coordinates of our theory are
noncommuting matrices in the group $U(N)$. 
In a region that the usual spacetime picture emerges, 
this theory  is described by the orbifold conformal field theory, 
essentially second quantized string theory in large N limit, 
contrary to the ordinary RNS
string theory which has a first quantized description. 

In the weak coupling limit that the SYM part must be considered and
the superconformal symmetry is preserved, a new additional dimension
appeared in the string spectrum, which is interpreted as compactified
eleven dimension in this paper. This additional degree of freedom is
interestingly described by the $S_N$-orbifold O(N) 
sigma model predicting that 
the size of this dimension increases in the ultraviolet limit 
and decreases in the infrared limit \c{on}. 
If a topological WZW term is included with the kinetic energy term, 
the size of the eleven dimension flows to some critical value in the
IR limit instead of flowing to zero size, where the purely bosonic
model is equivalent to a free fermion theory. But we don't ensure how
the WZW term can be naturally introduced in the NARNS string action. 
It will be interesting for these phenomena to give a natural M-theory
interpretation.  

Our matrix model is a worldsheet formulation compared to the
Green-Schwarz formulation of MST by DVV. 
While the MST of DVV is the type IIA string theory in
the Green-Schwarz light-cone formulation, it seems that the NARNS 
string theory provides a (non)-perturbative worldsheet matrix formulation of 
both type IIA and IIB superstring theory. The distinction of the type 
IIA and type IIB comes from
the GSO projection independently performed on the right and left
movers. The interaction of the NARNS string will be represented by a local
operator via a perturbation of the $S_N$-orbifold conformal field
theory \c{DVV}. Thus the identification of the vertex operator generating this
interaction would be important to obtain a picture of interacting
NARNS string. According to the recent argument \c{peri}, the
interaction generating the elementary joining and splitting of strings
may come from the holonomy of the gauge fields $v_m$ through the
monodromy of a twisted bundle. This implies that the gauge field on
the string worldsheet is crucial in the dynamics of the theory. 

Our theory also has two parameters like as MST \c{DVV}, 
$\alpha'$ and $g_s$, related with the
string tension $T=1/2\pi\alpha'$ and the Yang-Mills coupling constant 
$g_{YM}^{-2}=\alpha'g_s^2$. The coupling constant $g_s$ precisely 
plays the same role with the string coupling constant 
$\lambda_s=e^{-(\mbox{vev of dilaton})}$. If the string interactions
generating the elementary joining and splitting of strings 
are also correctly reproduced in our model, the coupling constant $g_s$ can
be directly related to the string coupling $\lambda_s$, since, in going
from the one-string to the two-string and vice versa, the genus of the
string worldsheet (Riemann surfaces) changes by one unit, 
so the vertex operator generating 
this interaction in our model must contain the factor 
$\lambda_s=e^{-(\mbox{vev of dilaton})}$. 

Compactification of the NARNS string theory means that some ``matrix'' 
string coordinates are defined over some compactified manifold. For
example, if we compactify a coordinate $X^I$ on a circle $S^1$ of radius
$R$, the components $X^I_{ij}$ (where the matrix indices $i$ and $j$
run from 1 to N) of the matrix $X^I$ can then be expanded in terms of 
the Fourier modes over $S^1$
\[X^I_{ij}(x)=\sum_n X_{ij,n}^I e^{inx/ {\hat R}},\;\;\;\;x\in S^1,
\;n\in {\bf Z} \]
where ${\hat R}= 1/2\pi R$. The $n$-th Fourier mode $X_{ij,n}^I$ 
exactly corresponds to the
matrix component $X^I_{0i,nj}$ of Ref.\c{taylor} which are based on
the D0-brane basis where $i$ and $j$ are labels of N D0-branes and the
coordinates $X^I_{mi,nj}$ are open string coordinates connecting the
$i$-th D0-brane within block $m \in {\bf Z}$ and $j$-th D0-brane within 
$n$-th block. Therefore the compactification scenario of \c{taylor} 
can be equally applied to the NARNS string theory. 

Although our matrix formulation of string theory is restricted only 
to the case of closed strings with $(1,1)$ worldsheet supersymmetry, 
e.g. type IIA and type IIB, 
it seems to be possible to extend to the cases of heterotic 
and type I strings. A naive conjecture for heterotic string and type I
string is following: The matrix fields of the heterotic string are
only 10-dimensional coordinates, 
$X_{R,L}^{0,\cdots,9}$ and $\psi_R^{0,\cdots,9}$ and the compactified
left-moving coordinates $X_L^{1,\cdots,16}$ still remain C-number
fields since they must identically generate the usual anomaly free 
gauge group $SO(32)$ and $E_8 \times E_8$ for the consitituent
strings. And, for the type I string, the identical $SO(32)$ Chan-Paton
factor is globally assigned for the consitituent strings at the
``boundary'' of the strings and ${\bf Z}_2$ modding of type IIB matrix string 
theory. However it is not obvious how this naive 
scenario consistently works. 

\section*{ACKNOWLEDGEMENTS}
This work was supported by the Korean Science and
Engineering Foundation through Center for Theoretical Physics, 
by the Korean Ministry of Education (BSRI-97-2414), and also by the
Sogang University Research Grant in 1997.

\appendix
\section{Conventions and Identities} 
The 2D bosonic metric is $\eta_{ab}$
\be
\l{ci1}
\eta^{ab}=\mbox{diag}(-1, +1)\;\; \mbox{and}\;\; 
\epsilon^{ab}=-\epsilon^{ba}; \;\; 
\epsilon^{01}=1.
\ee
The fermionic metric is $\epsilon^{\alpha\beta}$
\be
\l{ci2}
\epsilon^{\alpha\beta}=\epsilon_{\alpha\beta}=-\epsilon^{\beta\alpha};
\;\; \epsilon^{12}=1.
\ee
The spinors in this paper are Majorana in any case. 
Spin indices are raised and lowered by $\epsilon^{\alpha\beta}$
according to the rules
\be
\l{ci3}
\theta^\alpha = \epsilon^{\alpha\beta}\theta_\beta,\;\;\;
\psi_\alpha = \psi^\beta \epsilon_{\beta\alpha}, \;\;\;
{\bar \theta}\psi \equiv \theta^\alpha \psi_\alpha.
\ee

The Dirac $\gamma$-matrices are represented in the Majorana and Weyl
basis:
\bea
\l{ci4}
&&\gamma^a\gamma^b+\gamma^b\gamma^a=2\eta^{ab},\n\\
&&\gamma^0=\pmatrix{0 & 1 \cr -1 & 0},\;\;\;
\gamma^1=\pmatrix{0 & 1 \cr 1 & 0}, \;\;\;
\gamma_5=\gamma^0\gamma^1=\pmatrix{1 & 0 \cr 0 & -1}.
\eea
The index structure of the Dirac matrices is
$(\gamma^a)_\alpha \,^\beta$. 
The bilinear form ${\bar \psi}_1 \Gamma \psi_2$ for spinors 
${\bar \psi}_1$ and $\psi_2$ where $\Gamma$ is any combination of 
$\gamma$-matrices means that
\be
\l{ci5}
{\bar \psi}_1 \Gamma \psi_2 = \psi_1^\alpha
{(\Gamma)_\alpha}^\beta \psi_{2\beta}.
\ee 

The D-dimensional target space metric is $g^{IJ}$
\be
\l{ci6}
g^{IJ}=\mbox{diag}(-1, +1, \cdots, +1).
\ee

The Cartan-Killing metric for $U(N)$ Lie group in Eq.(\r{gaugep}) and
its Lie algebra are following
\be
\l{ci7}
Tr(T^r T^s)=\delta^{rs}, \;\;\; [T^r, T^s]=if^{rsp}T^p,
\ee
where $f^{rsp}$ is the structure constant of semi-simple Lie algebra
$su(N)$. 

Finally, we list the useful identies
\bea
\l{ci8}
&&({\bar \psi}_1 \psi_2)\psi_3=-\frac{1}{2}\sum_A
({\bar \psi}_1 \Gamma^A \psi_3)\Gamma_A\psi_2, 
\;\; \mbox{where}\;\;\Gamma^A=({\bf 1}, \gamma^a, \gamma_5), \n\\
&& \gamma^a\gamma^b\gamma_a=0,\\
&&{\bar \psi}_1\gamma^{a_1}\cdots \gamma^{a_n}\psi_2=
(-)^n{\bar \psi}_2\gamma^{a_n}\cdots \gamma^{a_1}\psi_1.\n
\eea

\section{Superspace of super-Yang-Mills theory coupled to 
2D supergravity}

Up to our knowledge, there is no reference constructing the superspace
formulation of 2D $N=1$ supergravity coupled to $N=1$
super-Yang-Mills theory. Thus, in this appendix, 
we will present the detailed solutions of the Binachi
identity (\r{gbi}) or (\r{gbic}) and discuss the supersymmetric
Yang-Mills theory coupled 2D supergravity in the context
of superspace formalism. 

In order to solve the super-Yang-Mills Bianchi identities we need the
results of the supergravity Bianchi identities (\r{bi}) subject to
the constraints (\r{ct}) due to Howe \c{howe}:
\bea
\l{t}
&&{T_{\alpha a}}^b=-{T_{a \alpha}}^b=0,\n\\
&&{T_{a\alpha}}^\beta=-{T_{\alpha a}}^\beta
=\frac{1}{4}{(\gamma_a)_\alpha}^\beta S,\\
&&{T_{ab}}^\alpha=\frac{i}{4}\epsilon_{ab}
(\gamma_5)^{\alpha\beta}D_\beta S,\n
\eea
where $S$ is the curvature superfield in the Eq.(\r{curv}).
We write down the Bianchi identities (\r{gbic}) in component form:
\be
\l{a1}
\Delta_{[a} F_{bc]}-{T_{[ab}}^\alpha F_{c]\alpha}=0,
\ee 
\be
\l{a2}
\Delta_{(\alpha} F_{\beta\gamma)}
-2i\gamma^a_{(\alpha\beta} F_{\gamma)a}=0,
\ee 
\be
\l{a3}
2\Delta_{[a} F_{b]\alpha}+\Delta_\alpha F_{ab}
-2{T_{\alpha[a}}^\beta F_{b]\beta}+{T_{ab}}^\beta F_{\alpha\beta}=0,
\ee
\be
\l{a4}
2\Delta_{(\alpha} F_{\beta)a}+\Delta_a F_{\alpha\beta}
+2{T_{a(\alpha}}^\gamma F_{\beta)\gamma}
-{T_{\alpha\beta}}^b F_{ab}=0. 
\ee
From Eq.(\r{a2}), one may obtain the relation
\bea
\l{a5}
F_{a\alpha}&=&-\frac{i}{2}{\gamma_a}^{\beta\gamma}
\Delta_\beta F_{\gamma\alpha}
-(\gamma^b\gamma_a)_\alpha\,^\beta F_{b\beta}\n\\
&=&-\frac{i}{6}{\gamma_a}^{\beta\gamma}
\Delta_\beta F_{\gamma\alpha}
+\frac{1}{3}(\gamma_a\gamma^b)_\alpha\,^\beta F_{b\beta}\\
&\equiv& {\gamma_{a\alpha}}^\beta W_\beta.\n
\eea
It follows from Eq.(\r{sfs}) and the 2D identity, 
$\gamma^a\gamma^b\gamma_a=0$ that $W_\alpha$ is of the form
\be
\l{a6}
W_\alpha=- \frac{i}{2}(\gamma_5)_\alpha\,^\beta
\Delta_\beta W(z).
\ee
Thus we obtain
\be
\l{a7}
F_{a\alpha}= \frac{i}{2}(\gamma_5 \gamma_a)_\alpha\,^\beta
\Delta_\beta W(z).
\ee
Using the above results, the Eq.(\r{a4}) leads to
\be
\l{a8}
F_{ab}=\epsilon_{ab}\frac{1}{4}(-\Delta^2 W + iWS).
\ee
It is also straightforward to obtain the following relations 
from the Eqs.(\r{a1}) and (\r{a2}):
\be
\l{a9}
\Delta_{(a}F_{bc)}=0,
\ee
\be
\l{a10}
\Delta_\alpha(\epsilon^{ab}F_{ab}+\frac{i}{2}WS)
+i{(\gamma^a)_\alpha}^\beta \Delta_a\Delta_\beta W=0.
\ee 
The first equation gives the usual Bianchi identity with respect
to the Yang-Mills field strengths $F_{ab}$.

In summary, the SYM Bianchi identities can be completely solved
together with the supertorsion components and the field strength
$F_{BC}$ can be represented by the scalar superfield $W$ which is
defined in terms of the spinor potential $A_\alpha$. 

For reader's  reference, we would like to present some sets of the
inverse superzweibein $E_A\,^M$ which are frequently used through this
paper 
\bea
\l{ine}
&&{E_\alpha}^m=i{(\gamma^m)_\alpha}^\mu \theta_{\mu}
-\frac{1}{4}{\bar \theta}\theta (\gamma^n\gamma^m)_{\alpha}\,^\mu
\chi_{n\mu},\\
&&{E_\alpha}^{\mu}={\delta_\alpha}^\mu 
+\frac{i}{2}\theta^{\nu}(\gamma^m)_{\nu\alpha}{\chi_m}^{\mu}
-\frac{i}{4}{\bar \theta}\theta \{(\gamma_5\gamma^m)_\alpha\,^\mu
\omega_m +\frac{i}{2}(\gamma^n\gamma^m)_{\alpha}\,^\nu\chi_{n\nu}
\;{\chi_m}^\mu+\frac{1}{2}{\delta_\alpha}^\mu A \} \n.
\eea

Under the gauge transformation (\r{gaugetr}) on $A_\alpha$, 
the component fields $\eta$ and $n$ shift by the gauge transformations
\bea
&&\delta_X \eta=i\zeta +ig_{YM}[\eta, \omega]\n\\
&&\delta_X n=i\rho +ig_{YM}[n, \omega]
+\frac{1}{2}({\bar \eta}\zeta+{\bar \zeta}\eta),\n
\eea
so that the fields $\eta$ and $n$ can be completely gauged away. 
In this Wess-Zumino (WZ) gauge, $\eta=n=0$, the gauge transformations 
(\r{gaugetr1}) of the component fields $v_m,\,\phi$, and $\lambda$ 
reduce to the ordinary gauge transformations:
\bea
&&\delta_\omega v_m=\nabla_m \omega,\n\\
&&\delta_\omega \phi =ig_{YM}[\phi, \omega],\n\\
&&\delta_\omega \lambda =ig_{YM}[\lambda, \omega].\n
\eea 

On the other hand, the supersymmetry transformation, (\r{strvec}), 
of component fields can be directly calculated by using the Eq.(\r{tpxi})
\bea
\l{a11}
&&\delta \eta=i\gamma^m \alpha\; g_m+i\gamma_5 \alpha\, \phi,\n\\
&&\delta n=-\frac{1}{2}{\bar \alpha}b
+\frac{1}{4}{\bar \alpha}\gamma^m \gamma^n\chi_m\, g_n
-\frac{1}{4}{\bar \alpha}\gamma^m\chi_m\, \phi,\n\\
&&\delta g_m=\frac{i}{2}{\bar \alpha}\gamma_mb+
\frac{i}{2}{\bar \alpha}\gamma^n\chi_m g_n
+\frac{i}{4}{\bar \alpha}\gamma_m \gamma^p \gamma^n\chi_p\, g_n 
-\frac{i}{4}{\bar \alpha}\gamma_5\gamma^n\gamma_m
\chi_n\, \phi, \\
&&\delta \phi=\frac{i}{2}{\bar \alpha}\gamma_5b+
+\frac{i}{4}{\bar \alpha}\gamma_5 \gamma^m \gamma^n\chi_m g_n
-\frac{i}{4}{\bar \alpha}\gamma^m
\chi_m\, \phi, \n\\
&&\delta b=\gamma^n\gamma^m\alpha\;{\cal D}_m g_n
+\gamma_5\gamma^m\alpha\;\partial_m \phi
-\frac{i}{2}{\bar \alpha}\gamma^m\chi_m \,\phi-\frac{i}{4}\gamma^n\chi_m\,
{\bar \alpha}\gamma^p\gamma^m\chi_p\,g_n \n\\
&&\;\;\;\;\;-\frac{i}{4}\gamma_5\chi_m\,
{\bar \alpha}\gamma^n\gamma^m\chi_n\,\phi
+\frac{1}{4}\gamma^m\alpha\, g_m A
-\frac{1}{4}\gamma_5\alpha\, \phi A,\n
\eea
where ${\cal D}_m g_n=e_{na}{\cal D}_m g^a$.
The above supersymmetry transformations violate the WZ gauge condition
$\eta=n=0$. We therefore need to make compensating gauge
transformation to maintain the WZ gauge. This is achieved by choosing
the gauge parameters as
\bea
\l{a12}
&&\zeta=-\gamma^m \alpha\; v_m-\gamma_5 \alpha\, \phi,\n\\
&& n =-\frac{i}{2}{\bar \alpha}\lambda
+\frac{i}{2}{\bar \alpha}\gamma^m \gamma^n\chi_m\, v_n
-\frac{i}{4}{\bar \alpha}\gamma^m\chi_m\, \phi.\n
\eea
From Eq.(\r{a11}), we can obtain the covariantized 
supersymmetry transformations (\r{stym}) for the Yang-Mills multiplet 
$(v_m,\,\phi,\,\lambda)$ by straightforward calculation. In the
calculation of the transformation $\delta \lambda$, we have used the
following useful identities with respect to the helicity decomposition of 
the gravitino $\chi_m$:
\bea
\l{igrno}
&&\chi_m= {\tilde \chi}_m+\gamma_m\chi, 
\;\;\;\gamma^m {\tilde \chi}_m=0,\n\\ 
&&\gamma_m{\tilde \chi}_n=\gamma_n{\tilde \chi}_m,
\;\;\;\gamma_5{\tilde \chi}_m=-e\epsilon_{mn}{\tilde \chi}^n,\\
&&2{\bar {\tilde \chi}}_m{\tilde \chi}_n=
g_{mn}{\bar {\tilde \chi}}\cdot{\tilde \chi},
\;\;\;{\bar {\tilde \chi}}_m \gamma_n {\tilde \chi}_p=0.\n
\eea

In the WZ gauge, the scalar field strength $W(z)$, (\r{sfse}), can be
calculated straightforwardly by using the Eqs.(\r{spinconn}), 
(\r{wzvm}), and (\r{ine}) and the result is
\bea
\l{W}
&&W(z)=-2i\phi +2{\bar \theta}\gamma_5 \lambda 
-\frac{1}{2}{\bar \theta}\gamma^m \chi_m \phi\n\\
&&\;\;\;\;\;\;\;\;\;\;-\frac{1}{2}{\bar \theta}\theta
\{\frac{\epsilon^{mn}}{e}v_{mn}
+i{\bar \lambda}\gamma_5\gamma^m\chi_m
-\frac{i}{4}{\bar \chi}_n\gamma^m\gamma^n\chi_m\phi+A\phi\}.
\eea
It is now straightforward 
although somewhat tedious to calculate 
$\nabla_\alpha W, \; \nabla_\alpha W\,\nabla^\alpha W$ and 
$E\,\nabla_\alpha W\,\nabla^\alpha W$ from the Eq.(\r{W}) 
to obtain the super-Yang-Mills
action (\r{sym}).

\end{document}